\documentclass[reprint,aps,pra]{revtex4-1}

\usepackage{graphicx}
\usepackage{dcolumn}
\usepackage{bm}
\usepackage{amsmath}
\usepackage{amsthm}
\usepackage{amssymb}
\usepackage{float}
\usepackage{hyperref}
\usepackage{color}
\usepackage{epstopdf}
\usepackage{cleveref}
\usepackage[svgnames]{xcolor}
\usepackage{multirow}
\hypersetup{hidelinks,colorlinks=true,allcolors=DarkBlue}

\begin{document}

\preprint{APS/123-QED}

\title{Influence of the laser frequency drift \\ in phase-sensitive optical time-domain reflectometry}
\author{A.A. Zhirnov$^{1}$}
\author{K.V. Stepanov$^{1}$}
\author{A.O. Chernutsky$^{1}$}
\author{A.K. Fedorov$^{2}$}
\author{E.T. Nesterov$^{1}$}
\author{C. Svelto$^{3}$}
\author{A.B. Pnev$^{1}$}
\author{V.E. Karasik$^{1}$}
\affiliation
{
	\mbox{$^{1}$Bauman Moscow State Technical University, Moscow 105005, Russia}
	\mbox{$^{2}$Russian Quantum Center, Skolkovo, Moscow 143025, Russia}
	\mbox{$^{3}$Dipartimento di Elettronica e Informazione del Politecnico di Milano, Milano 20133, Italy}
}

\begin{abstract}
The influence of the laser frequency drift on the operation of phase-sensitive optical time domain ref lectometry ($\Phi$-OTDR) systems is considered. 
Theoretical results based on a new numerical $\Phi$-OTDR model demonstrating the influence of the laser frequency instability on a signal are reported. 
This model is verified based on experimental data. It has been used to calculate the signal-to-noise ratio (SNR) of the system for different parameters of the laser source stability. 
As a result, quantitative requirements for lasers used in $\Phi$-OTDR systems are formulated.
\end{abstract}

\date{\today}
                    
\maketitle

\section{Introduction}

Distributed vibration sensors based on phase-sensitive optical time domain reflectometry are highly promising for remote monitoring of extended objects, such as bridges, roads, and pipelines~\cite{Bao2012}. 
In contrast to conventional OTDR sensors, a narrow-band probe signal makes it possible to detect perturbations based on their influence on the signals (Fig.~1), 
which are formed due to phases of backscattered waves~\cite{Taylor1993,Park1998,Choi2003}. 
In recent decades, $\Phi$-OTDR sensor systems have attracted a significant deal of interest~\cite{Taylor1993,Park1998,Choi2003,Juarez2005,Rao2009,Lu2010,Martins2013,Martins20132,Pnev2015,Alekseev2016,Nikitin2016,Alekseev2015}.
Modern developments in production of fiber-optic components have reduced the production cost of such sensor systems significantly and increased their quality, which expands the range of their applications.

\begin{figure}
	\includegraphics[width=.85\linewidth]{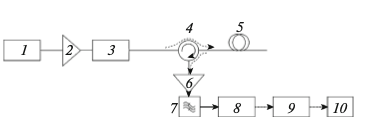}
	\vskip -5mm
	\caption
	{Schematic of the phase-sensitive optical time domain reflectometer.}
	\label{fig:scheme}
\end{figure}

The principle of operation of the system can be described by the example of the schematic shown in Fig.~\ref{fig:scheme}. 
The radiation from narrow-band source 1 is amplified to a desired power level in booster 2 and passes to the pulsed regime in acoustooptic modulator 3. 
Then it is fed to sensing fiber 5 via circulator 4. The backscattered radiation is first amplified in preamplifier 6. 
The spontaneous radiation is rejected by optical filter 7. 
Then the signal is recorded by photodetector 8, digitized by ADC 9, and processed by PC 10. 
A response of this system appears when the signal begins to intensely fluctuate in some stage with respect to its level during the recent time interval.

Phase optical time domain reflectometry is implemented only when the radiation source coherence length is not smaller than the pulse width~\cite{Taylor1993,Park1998,Choi2003}.  
Thus, the laser is the decisive component of the system. 
In general, the quality of sensor systems for remote monitoring of extended objects is determined 
by their ability to detect perturbations caused by some activity near the fiber against the background of both external noises (for example, seismic) and intrinsic noises of the system. 
To analyze their influence, Ph-OTDR sensors were also investigated based on numerical simulation~\cite{Pnev2015,Alekseev2011,Gabai2016,Alekseev2012,Li2014,Liokumovich2015,Zhong2014,Pnev20152}.

In some studies~\cite{Martins2013,Zhong2014}, the role of the laser source was investigated. 
In particular, it was demonstrated that the source frequency drift may reduce the SNR and even cause spurious responses. 
However, no quantitative criteria of the frequency stability of laser sources for Ph-OTDR sensors have been proposed to date. 
One reason for this is the presence of different laser types for Ph-OTDR with specific features. 
The fiber, solid-state, or semiconductor lasers in use may have fluctuations at different frequencies. 
The power spectral density of frequency noise may serve as a more general characteristic. In this study, we present the description of the general view of this characteristic, 
which will make it possible to solve in the first approximation the problem of its different shape for different sources.

In this paper, we report the results of studying the influence of frequency instability of laser sources on the signal of Ph-OTDR systems using numerical simulation and experimental investigation. 
First of all, we propose a new numerical model considering frequency fluctuations in time. 
To describe it, we present the general parameterized form of the power spectral density of frequency fluctuations as the upper limit of the current model. 
This assumption is verified and justified by experimental data for three different lasers. 
Furthermore, we calculate the SNR under similar conditions for each laser based on the generated arrays of Ph-OTDR signals from different sources set by the frequency-stability parameters. 
This will allow us to choose the range of parameters that satisfy the high-quality operation of the sensor system. 
Thus, the requirements for frequency fluctuations for sources in Ph-OTDR systems, which are aimed at the maximization of true detection of events, can be formulated numerically.

\section{Numerical simulation}

The signal of a phase optical time domain reflectometer is composed of backscattered waves with a random phase distribution. 
This signal nature allows for only statistical description of this system. To develop the model, we should describe the parameters of its elementary particle (one scattering center). 
The model is constructed based on the fact that scattering from all centers is a random complex signal~\cite{Goodman}:
\begin{align}
\begin{split}
	p(a)&=\left\{\begin{array}{ll}
		\frac{a}{\sigma^2}\exp\left[-\frac{a}{2\sigma^2}\right],&\mbox{ if }a>0,\\
		0,&\mbox{ if } a<0.
	\end{array}\right. \\
	p(\Theta)&=\left\{\begin{array}{ll}
		\frac{1}{2\pi},&\mbox{ if }\Theta\in\left({-\pi,\pi}\right],\\
		0,&\mbox{ otherwise}.
	\end{array}\right.
\end{split}
\label{eq:Rayleigh}
\end{align}
Here $p(a)$ is the scattering amplitude distribution density and $p(\Theta)$ is the phase distribution density of scattered waves.

One should keep in mind that the scattering amplitude also depends on distance to the $n$th scattering center $L_n$: 
the value of attenuation $\exp(2\alpha L_n)$ should be considered, where $\alpha$ (in dB/km) is the attenuation coefficient in the fiber and 2 is a factor taking into account light propagation to the scattering center and backward. 
The size of inhomogeneities that are relevant for the Rayleigh scattering is about one-tenth of the wavelength; correspondingly, this value is about $d_c=150$ nm for a device operating at $\lambda\approx1550$ nm. 
The distribution in the mode cross section can be disregarded, because phases of all backscattered waves (from these centers) are identical. 
Accordingly, for complete simulation of a sensor with $L_c=50$ km, one must calculate roughly the following number of elements:
\begin{equation}
	N_p=\frac{L_c}{d_c}=\frac{50\times10^3}{150\times10^{-9}}=3\times10^{11} \mbox{ elements}.
\end{equation}
Because of limited computational resources, we will find the ways to reduce the number of objects for model calculations. 
Since the damping coefficient $\alpha$ is about $0.2$ dB/km for the operating wavelength of 1550 nm in modern types of fibers,
the related differences between the scattering centers within one pulse (up to 100 m) can be considered negligible. 
In addition, there is no need to simulate the whole sensor, because the event is generally localized at a portion of not more than 100 m; 
i.e., it is sufficient to set the sensor length to be $L_{ce}=500$ m and provide the corresponding (for the portion under study) damping. 
We also suggest preliminary summation of waves from a portion that is smaller than the spatial instrumental resolution and the wavelength in soil. 
Scattering amplitude distribution from these portions de also has the shape of the Rayleigh distribution
\begin{equation}
	N_p=\frac{L_{ce}}{d_e}=\frac{500}{0.05}=10^{4} \mbox{ elements}.
\end{equation}
This value of the array allows one to perform simu- lation without deterioration of the quality of results. 
We Note that, when carrying out numerical experiments, it is interesting to trace the time evolution of the scat- tered signal; 
i.e., it is necessary to retain $10^4$ elements for at least $60\times10^3$ steps, which is equivalent to a 1-min signal of the phase optical time domain reflectometer recorded with a frequency of 1 kHz.

To obtain information about the signal, one should set the following main parameters of the system:

(i) wavelength $\lambda_t$ or frequency $\nu_t$ of the laser source for each instant of feeding a probe pulse to the sensor;

(ii) amplitude $A_n$ of the $n$th scattering center (it is constant during all evolutions of the model); and

(iii) instead of setting phase of each center $\varphi_{n,t}$ one should set distance from the laser source (or any other conditional point of the sensor origin) to this center $L_n$; 
the phase is calculated from the following relation: 
\begin{equation}\label{eq:4}
	\varphi_{n,t}=\frac{2\pi}{\lambda_t}L_n=\frac{2pin_g\nu_t}{c}L_n,
\end{equation}
where $n_g$ the group effective refractive index of the fiber core.

A simple and efficient way is to set a uniform increase in the distance between the scattering centers and the origin.
A set of the above-mentioned parameters will yield a value of the complex signal from each specified portion of scattering:
\begin{equation}
	s_{t,n}=A_ne^{j\varphi_{n,t}}.
\end{equation}

Furthermore, we should obtain the signal that is detected by the ADC at each instant when recording one reflectogram. 
This is a squared magnitude of the sum with allowance for the phases of signals from the scattering centers in a half a length of the optical pulse
\begin{equation}
	S_{t,ni}=\left|{\sum_{n=n_i}^{n_i+n_{{\rm imp}/2}}{s_{t,n}}}\right|,
\end{equation}
where $n_i$ is the first scattering center at a current ADC indication and $n_{{\rm imp}/2}$ is the number of centers scattering fit into a half-width of the probe pulse.

In the model, the shift of the probe pulse is determined in terms of the number of elements:
\begin{equation}
	\Delta{n_s}=\frac{t_{\rm ADC}c}{2n_ed_e}.
\end{equation}

In this case, one reflectogram of $K$ indications at the ADC $(K=L_{ce}/(d_e\Delta{n_s}))$ can be obtained from the following expression:
\begin{equation}
	R_{{\rm ind}, t}=S_{t,ni}|_{i=({\rm ind}-1)\Delta{n_s}+1}^{K\Delta{n_s}+1},
\end{equation}
where ${\rm ind}$ is the reflectogram indication number.

Similarly, using special functions,
\begin{equation}
	\!\!\!\!\!\!R_{{\rm ind}, t}={\rm comb}\left({\frac{i}{\frac{t_{\rm ADC}c}{2n_ed_e}}}\right)\left(S_{t,ni}|_{i=1}^{L_{ce}/d_e}\otimes{{\rm comb}\frac{i}{\frac{\tau_{\rm imp}c}{2n_ed_e}}}\right)
\end{equation}
This formula is useful, because it can be set faster for calculations and allows one to simulate deviations of the pulse shape from the ideal one with rect replaced by a numerically specified geometry.

\begin{figure}
	\includegraphics[width=\linewidth]{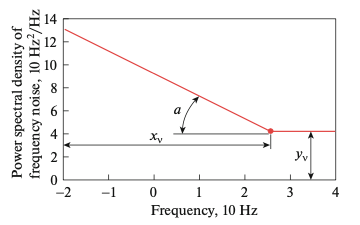}
	\vskip -5mm
	\caption
	{Algorithm of determining parameters for the laser simulation.}
	\label{fig:algorithm}
\end{figure}

The final formation of one reflectogram in the form intended for demonstration on the screen requires consideration of the bandwidth of the electric amplifier of the photodiode detector. 
Since the high-est-frequency signal (up to 100 MHz) in the system contains a large fraction of noise, it is generally filtered to the range from 0 to 5 MHz
\begin{equation}
	Rf_{{\rm ind},t}=R_{{\rm ind},t}\otimes{h_{pr}(i)}+N_{{\rm ind},t},
\end{equation}
where $h_{pr}(i)$ is the pulsed function of the electric amplifier and $N_{{\rm ind},t}$ is the remaining physical (from vibrations of the scattering centers), optical (from the preamplifier), and electric noises of the system.
If there is no external signal effect, the changes are affected by two main factors:
(i) time fluctuations of central laser wavelength $\nu_t$ and (ii) vibrations of the scattering centers under temperature and random external mechanical actions on the sensing fiber.

In most cases, the change in the signal is due to specifically the first reason, provided that the fiber is not placed in a region with strong constant noise effects.

Producers of laser sources most often offer the plot of power spectral density of frequency fluctuations $S_\nu$ as a characteristic of the central wavelength stability.
It is specified in the range from $1/T=0.01$ Hz (for setting in the time interval of $T=100$ s in order to eliminate edge effects) to reflectogram repetition rate $\nu_p=1$ kHz.

Generally, plot envelope $\hat{S}_\nu$ is generally provided by a manufacturer; to find the realistic time dependence, one should multiply this characteristic by the normalized spectral density of a random uniform signal:
\begin{equation}
	{S}_\nu=\hat{S}_\nu\left|{F\left\{{\rm rand}(1)\right\}}\right|\frac{1}{\nu_p^2T},
\end{equation}
where $F\left\{\dots\right\}$ is the Fourier transform. 

Then, the time dependence of the frequency can be obtained using the following transformation:
\begin{equation}
	\nu_t=\nu_0+\left|{F^{-1}\left\{{\sqrt{{S}_\nu \nu_p^2T}}\right\}}\right|,
\end{equation}
where $\nu_0$ is the laser frequency at the initial instant and $F^{-1}\left\{\dots\right\}$ is the inverse Fourier transform. 

It is convenient to set the general view of the envelope in the form of a plot shown in Fig.~\ref{fig:algorithm}.
It consists of linearly decreasing and constant parts. In this form, the envelope can be conveniently set using three parameters: the slope of the decreasing part and the frequency and value of the spectral density at the inflection point.

This model allows one to obtain the simulated signal from the phase-sensitive optical time domain reflectometer using a certain laser with a specified characteristic, which exhibits no external effects. 
To set the latter, one should specify deformation of the sensing fiber $\Delta{L_{n,t}}$ under the action of an external vibrational signal. 
Transfer of mechanical vibrations from environment to the fiber cable is a complex and multiparameter process and depends on many parameters (e.g., density, humidity, viscosity, temperature, granularity, and wetting for soil). 
In addition, the transfer of the effect through the fiber shell will be different in each case (e.g., cable armoring). 
This question is not considered in detail due to the high complexity, and the process of external effect transfer is interpreted according to the following algorithm:
(i) the effect exerted at some point causes a pulsed single displacement of the scattering centers by $\Delta{L_{\rm max}}$;
(ii) after receiving the initial deformation, the scattering centers begin to vibrate with period $P(s)$ determined by the properties of environment and fiber $s$ damping with time constant $r_t(s)$; and
(iii) vibrations also propagate from the site of initial effect with a velocity $v(s)$ damping with the distance constant $r_n(s)$.

\begin{widetext}

Then the effect occurred at instant $t_c$ in portion $n_c$ of the sensor can be set by the expression
\begin{equation}\label{eq:12}
	\Delta{L_{n,t}(\Delta{L_{\rm max}},s,n_c,t_c)}=
	\begin{cases}
	\Delta{L_{\rm max}}\exp[-\frac{|n-n_c|}{\tau_n(s)}]\exp[-\frac{|t-t_c|}{\tau_n(s)}]\sin[\frac{t-t_c-\frac{d_e|n-n_c|}{v(s)}}{P(s)}] &t \geq t_c\\
	0 &\text{otherwise}.
	\end{cases}
\end{equation}

\end{widetext}

This parameter is an additional term in the position of the scattering centers $L_n$ in expression~(\ref{eq:4}). 
Specifying a set of effects in accordance with expression~(\ref{eq:12}), we can obtain simulated signals from the sensor system with external effects.

\section{Experimental part}

\begin{table*}[t]
\begin{tabular}{|c|c|c|c|c|}
	\hline
	\multirow{2}{*}{Laser} & \multicolumn{2}{c|}{Standard deviation of the long portion} & \multicolumn{2}{c|}{Standard deviation of the short portion} \\ \cline{2-5} 
                       & Experiment                     & Model                      & Experiment                      & Model                      \\ \hline
	OE                     & 0.1336                         & 0.1188                     & 0.1226                          & 0.1175                     \\ \hline
	DL                     & 0.1150                         & 0.1043                     & 0.0297                          & 0.0313                     \\ \hline
	RIO                    & 0.0431                         & 0.0424                     & 0.0345                          & 0.0360                     \\ \hline
	\end{tabular}
	\caption{Comparison of the standard deviations for the experimental and simulation data.}
	\label{tbl:1}
\end{table*}

\begin{figure}
	\includegraphics[width=\linewidth]{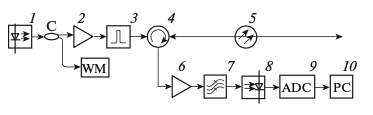}
	\vskip -5mm
	\caption
	{Supplemented schematic of the phase-sensitive optical time domain reflectometer.}
	\label{fig:algorithm}
\end{figure}

To verify the proposed model and its results, we compare the numerical and experimental data for three laser sources.
To carry out this experiment, the schematic of the phase optical time domain reflectometer was supplemented with a site providing simultaneous (with detection of the signal) measurement of the source wavelength. 
To this end, some laser radiation was fed to a wavelength meter (WM) via splitter $C$.

\begin{figure*}[htbp]
\includegraphics[width=2\columnwidth]{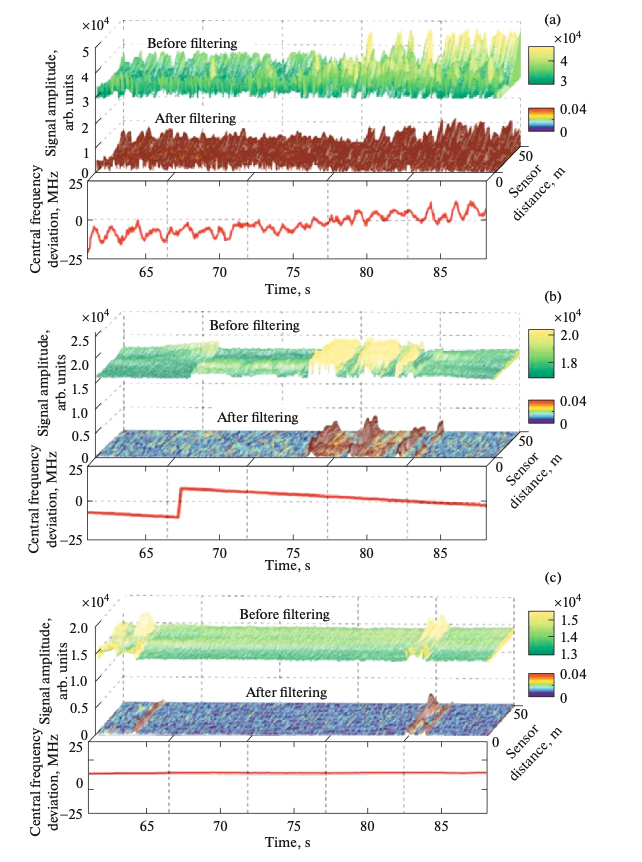}
\caption
{Signal samples for (a) OE Waves, (b) DenseLight, and (c) RIO recorded simultaneously with the detection of the wave- length source. These signals filtered by a bandpass filter from 8 to 40 Hz are also presented.}
\label{fig:samples}
\end{figure*}

\begin{figure*}[htbp]
\includegraphics[width=2\columnwidth]{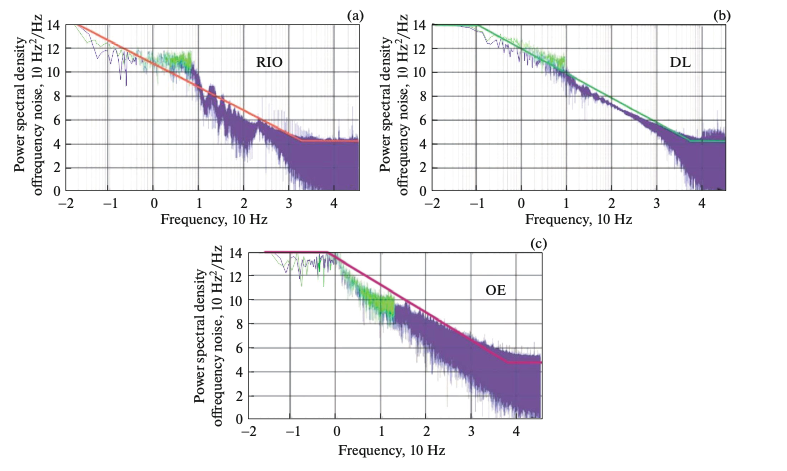}
\caption
{Power spectral densities of frequency fluctuations of the (a) RIO, (b) DenseLight, and (c) OE Waves lasers.}
\label{fig:power}
\end{figure*}

In this schematic (Fig. 3), the shown reflectometer signal (Fig.~\ref{fig:samples}) and the power spectral density of frequency fluctuations for each source (Fig.~\ref{fig:power}) can be simultaneously measured. 
For clarity, a change in the laser frequency in a time interval, in which the reflectometer signal was detected, is also shown in Fig.~\ref{fig:samples}.

\begin{figure*}[htbp]
\includegraphics[width=2\columnwidth]{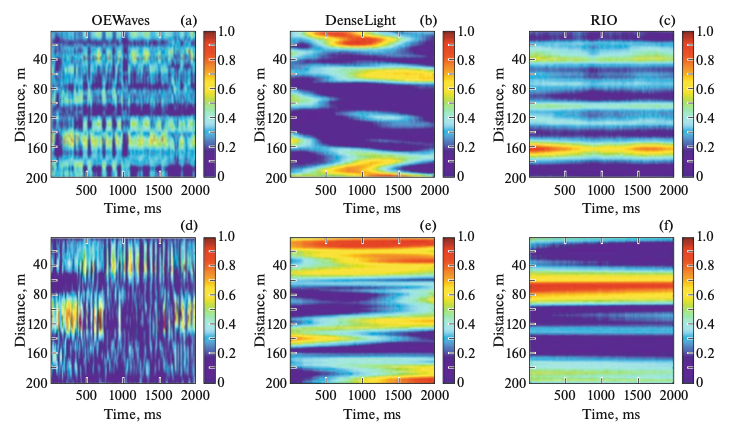}
\caption
{Signals of the phase-sensitive optical time domain reflectometer: (a--c) experimental data and (d--f) numerical simulation data.}
\label{fig:signals}
\end{figure*}

Based on the measured spectral densities of the sources, the signals of reflectometers with these sources were simulated according to the proposed model (without external effects). 
The results of simulating the space--time field distribution $I(l, t)$ and experimental data can be visually compared in Fig.~\ref{fig:signals} and using the standard deviation values in the short (50 ms) and long (1000 ms) intervals given in Table~\ref{tbl:1}. 
One can see that the results obtained are in good correspondence, which is indicative of the correctness of the developed model.

\section{Computational part}

\begin{figure*}[htbp]
\includegraphics[width=2\columnwidth]{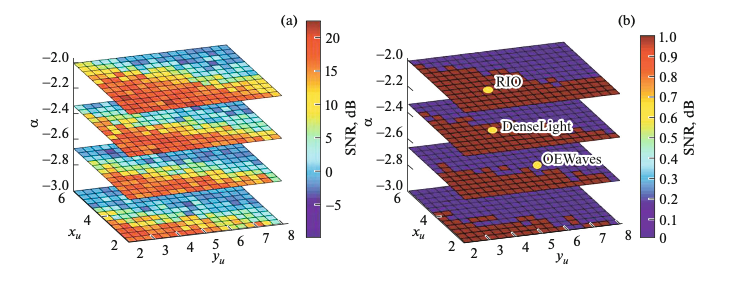}
\caption
{(a) SNR map and (b) regions of lasers providing SNR$>10$.}
\label{fig:SNR}
\end{figure*}

To determine the quantitive characteristics of laser sources providing the operating capacity of the phase-sensitive optical time domain reflectometer, 
we simulate a signal from the system with lasers having different characteristics.

(i) The plot consists of only the linearly decreasing part and uniform noise background. Position of the inflection point $(x_\nu, y_\nu)$ is set for the simulation.

(ii) The linearly decreasing part of the plot has slope $\alpha$ from $-20$ to $-30$ dB/order.

(iii) The SNR was determined using the following algorithm. 
A sinusoidal signal was generated at the first 60 m of the sensor. 
Since the generated intensity of one count was from 0 to 40 arb. units for the specified pulse width, a portion with the average intensity that is closest to 20 arb. units was chosen in the first few meters of the sensor.
This value makes it possible to estimate the signal in a portion with the mean sensitivity, avoiding portions with a too low or too high sensitivity. 
The signal standard deviation in time, normalized to the value averaged over $N$ reflectograms, is calculated for this portion,
\begin{equation}
\begin{split}
	{\rm STD_{sin}}&=\frac{1}{J_{\rm sig}}\sqrt{\frac{\sum_i^N{(J_{i,j20}-J)^2}}{n}}, \\
	\overline{J_{\rm sig}}&=\sum_i^N{J_i,j20}.
\end{split}
\end{equation}

For the region, in which the effect was not generated, the normalized standard deviation was determined for portions with the minimum and maximum averages; the final value was found as their arithmetic mean. 
The quantity obtained is considered to be the noise in this simulation,
\begin{equation}
\begin{split}
	{\rm STD_{noise}}&=\frac{1}{2J}\sqrt{\frac{\sum_i^N{(J_{i,j\,{\rm min}}-\overline{J_{\rm sig}})^2}}{n}}+, \\
	&\frac{1}{2J}\sqrt{\frac{\sum_i^N{(J_{i,j\,{\rm max}}-\overline{J_{\rm max}})^2}}{n}} \\ 
	&\overline{J_{\rm min},{\rm max}}=\sum_i^N{J_i,j,{\rm min},{\rm max}}.
\end{split}
\end{equation}

Thus, the SNR is determined by the ratio of the above parameters:
\begin{equation}
	{\rm SNR}=\frac{{\rm STD_{sin}}}{{\rm STD_{noise}}}.
\end{equation}

The simulation results are presented in Fig.~\ref{fig:SNR}a as a color map. The region of sources, providing the SNR of more than 10 in the system signal, is colored in Fig.~\ref{fig:SNR}b.

The data obtained show that two of three lasers used in the experiment make it possible to obtain a high-quality signal. 
These results coincide with the visual representation in Fig.~\ref{fig:samples}, where the signals after filtering in the range from 8 to 40 Hz are shown. 
In the first two panels, the signal exhibits responses of the sensor to impacts, whereas in the third panel only the constantly fluctuating signal remains, which makes it impossible to distinguish events.

Proceeding from Fig.~\ref{fig:SNR}b, one can state that, to obtain a high-quality signal in the reflectometer, a laser with a power spectral density of frequency fluctuations not exceeding $10^3$ Hz$^2$/Hz at a frequency of 1 kHz should be applied. 
However, for more accurate determination of the SNR for the laser source under study, the best solution entails substitution of its spectral density of frequency fluctuations into the model and subsequent calculation of the SNR.

\section{Conclusions}
A mathematical model for the formation of a signal of a phase optical time domain reflectometer was developed and verified by consistency with the experimental data. 
It takes into account frequency fluctuations of the laser source. 
Based on this model, quantitative criteria to the sources providing a high-quality signal of a phase optical time domain reflectometer were proposed by a series of numerical calculations.

{\bf Acknowledgments}.
A theoretical and experimental study of the stability of a laser source was performed by A.A. Zhirnov and K.V. Stepanov and funded by RFBR according to the research project No.18-32-00688. 
A.K. Fedorov was funded by the RFBR according to the research project No. 18-37-20033.

\end{document}